\documentclass[showpacs,twocolumn,preprintnumbers,amsmath,amssymb,amsbsy,amsfonts,prb,floatfix, superscriptaddress]{revtex4}

\usepackage{graphicx, color,rotating, textcomp} 
\usepackage{dcolumn} 
\usepackage{bm} 

\begin{document}

\title{Inverse proximity effect and influence of disorder on triplet supercurrents in strongly spin-polarized ferromagnets}

\author{Roland~Grein}
\affiliation{Institut f\"ur Theoretische Festk\"orperphysik
and DFG-Center for Functional Nanostructures,
Karlsruhe Institute of Technology, D-76128 Karlsruhe, Germany}
\affiliation{SEPnet and Hubbard Theory Consortium, Department of Physics, Royal Holloway, University of London, Egham, Surrey TW20 0EX, United Kingdom}
\author{Tomas~L\"ofwander}
\affiliation{Department of Microtechnology and Nanoscience MC2,
Chalmers University of Technology, S-412 96 G\"oteborg, Sweden}
\author{Matthias~Eschrig}
\affiliation{SEPnet and Hubbard Theory Consortium, Department of Physics, Royal Holloway, University of London, Egham, Surrey TW20 0EX, United Kingdom}
\date{\today}

\begin{abstract}
We discuss the Josephson effect in strongly spin-polarized ferromagnets where triplet correlations are induced by means of spin-active interface scattering, extending our earlier work [R. Grein et al., Phys. Rev. Lett. \textbf{102}, 227005 (2009)] by including impurity scattering in the ferromagnetic bulk and the inverse proximity effect in a fully self-consistent way. Our quasiclassical approach accounts for the differences of Fermi momenta and Fermi velocities between the two spin bands of the ferromagnet, and thereby overcomes an important short-coming of previous work within the framework of Usadel theory. We show that non-magnetic disorder in conjunction with spin-dependent Fermi velocities may induce a reversal of the spin-current as a function of temperature.\end{abstract}

\maketitle

\newcommand{\pF}{\vec{p}_{\rm F}}
\newcommand{\vphi}{\varphi}
\newcommand{\eps}{\varepsilon}
\newcommand{\ud}{\uparrow,\downarrow}
\renewcommand{\u}{\uparrow}
\renewcommand{\d}{\downarrow}
\newcommand{\ket}[1]{| {#1}\rangle}
\newcommand{\bra}[1]{\langle {#1}|}
\newcommand{\barlambda}{{\lambda \!\!\!^{-}\,\!}}
\newcommand{\blFe}{{\lambda \!\!\!^{-}\,\!}_{\mathrm{F}1}}
\newcommand{\blFeta}{{\lambda \!\!\!^{-}\,\!}_{\mathrm{F}\eta}}
\newcommand{\blF}{{\lambda \!\!\!^{-}\,\!}_{\mathrm{F}}}
\newcommand{\blJ}{{\lambda \!\!\!^{-}\,\!}_{J}}
\newcommand{\EF}{E_{\mathrm{F}}}
\newcommand{\vpfe}{\vec{p}_{\mathrm{F}1}}
\newcommand{\vpfz}{\vec{p}_{\mathrm{F}2}}
\newcommand{\vpfd}{\vec{p}_{\mathrm{F}3}}
\newcommand{\vpfeta}{\vec{p}_{\mathrm{F}\eta }}
\newcommand{\vvfe}{\vec{v}_{\mathrm{F}1}}
\newcommand{\vvfz}{\vec{v}_{\mathrm{F}2}}
\newcommand{\vvfd}{\vec{v}_{\mathrm{F}3}}
\newcommand{\vvfzd}{\vec{v}_{\mathrm{F}2,3}}
\newcommand{\vvfeta}{\vec{v}_{\mathrm{F}\eta }}
\newcommand{\pfe}{p_{\mathrm{F}1}}
\newcommand{\pfz}{p_{\mathrm{F}2}}
\newcommand{\pfd}{p_{\mathrm{F}3}}
\newcommand{\pfeta}{p_{\mathrm{F}\eta }}
\newcommand{\vfe}{v_{\mathrm{F}1}}
\newcommand{\Nfe}{N_{\mathrm{F}1}}
\newcommand{\vfz}{v_{\mathrm{F}2}}
\newcommand{\Nfz}{N_{\mathrm{F}2}}
\newcommand{\vfd}{v_{\mathrm{F}3}}
\newcommand{\Nfd}{N_{\mathrm{F}3}}
\newcommand{\vfzd}{v_{\mathrm{F}2,3}}
\newcommand{\vfeta}{v_{\mathrm{F}\eta}}
\newcommand{\JFM}{J_{\mathrm{FM}}}
\newcommand{\vJFM}{\vec{J}_{\mathrm{FM}}}
\newcommand{\JI}{J_{\mathrm{I}}}
\newcommand{\vJI}{\vec{J}_{\mathrm{I}}}
\newcommand{\VI}{V_{\mathrm{I}}}
\newcommand{\gr}{\gamma^R}
\newcommand{\ga}{\gamma^A}
\newcommand{\grt}{\tilde{\gamma}^R}
\newcommand{\gat}{\tilde{\gamma}^A}
\newcommand{\gra}{\gamma^{R,A}}
\newcommand{\grat}{\tilde{\gamma}^{R,A}}
\newcommand{\Rs}{{R}_1}
\newcommand{\Tsn}{{T}_{sn}}
\newcommand{\Tns}{{T}_{ns}}
\newcommand{\Rn}{{R}_n}
\newcommand{\ru}{r_{2}}
\newcommand{\rd}{r_{3}}
\newcommand{\Tsu}{{T}_{12}}
\newcommand{\Tsd}{{T}_{13}}
\newcommand{\Tus}{{T}_{21}}
\newcommand{\Tds}{{T}_{31}}
\newcommand{\rud}{r_{23}}
\newcommand{\rdu}{r_{32}}
\newcommand{\vecb}[1]{\mathbf{#1}}
\newcommand{\Tc}{T_\mathrm{c}}

\section{Introduction}

Ferromagnet-superconductor hybrid systems are currently subject to intense research activity, as they were conjectured to host triplet superconductivity induced by the proximity effect\cite{bergeret05, buzdin05, eschrig07, lofwander10,eschrig11}. While the first successful experiments on these structures found evidence for so-called $S_z=0$ triplet correlations\cite{ryazanov01, kontos02, blum02}, whose hallmark are $0-\pi$-oscillations \cite{buzdin05} of the critical current in Josephson junction devices, the existence of equal spin triplet pairing is currently in the focus of attention\cite{bergeret05, eschrig11}. So far, the main experimental prove of such triplet correlations is based on the Josephson effect and on phase coherent electron transport in proximity structures \cite{keizer06, sosnin06,robinson10a, robinson10b, khaire10, khasawneh11, anwar10, sprungmann10}. After two first confirmations of long-range triplet amplitudes
in 2006, \cite{keizer06, sosnin06} an impressive series of affirmative experimental results was published in 2010.\cite{robinson10a, robinson10b, khaire10, anwar10, sprungmann10} The systems under investigation varied largely in terms of materials and fabrication, the common idea being that a breaking of spin-rotation symmetry around the bulk magnetization axis must somehow be enforced at the superconductor (SC)-ferromagnet (FM) contacts - which is in line with the general theoretical concepts behind this effect\cite{bergeret01,bergeret05,volkov03,eschrig03, houzet07, eschrig08}. In this article, we elaborate on our earlier work concerning long-range triplet supercurrents\cite{grein09}, where we used a recent extension of the quasiclassical Green function technique\cite{eschrig09}, that allows us to consider ferromagnets whose exchange splitting $J$ is of the same order of magnitude as the Fermi energy $E_\mathrm{F}$ consistently within quasiclassical theory\cite{grein09, grein10}. In the following we include additional relevant effects like the suppression of the superconducting energy gap in proximity to the interfaces, the induced magnetization in the superconductor close to the ferromagnet, and disorder in the FM bulk. These calculations are performed numerically and cover all ranges of junction length, interface transparency, and impurity scattering. In the appendix, we provide detailed discussions of the intricacies involved in solving the self-consistency equation in quasiclassical theory with spin-active boundary conditions. In particular, we point out a previously not discussed technical problem with the self consistent solution of the order parameter that arises for quasiclassical transport equations with boundary conditions where the quasiclassical propagator undergoes a jump discontinuity during reflection from the interface. 

With regard to the inverse proximity effect, we show that while the suppression of the gap function in the SC-electrodes is indeed substantial for highly transparent junctions, this does not imply that higher harmonic contributions to the current-phase relation (CPR) are necessarily suppressed. As expected, impurity scattering reduces the Josephson current and in particular higher harmonic contributions to the CPR. We also find that there may be a reversal of the Josephson spin-current as a function of temperature if impurity scattering is sufficiently strong.
\begin{figure}
\includegraphics[width=\columnwidth]{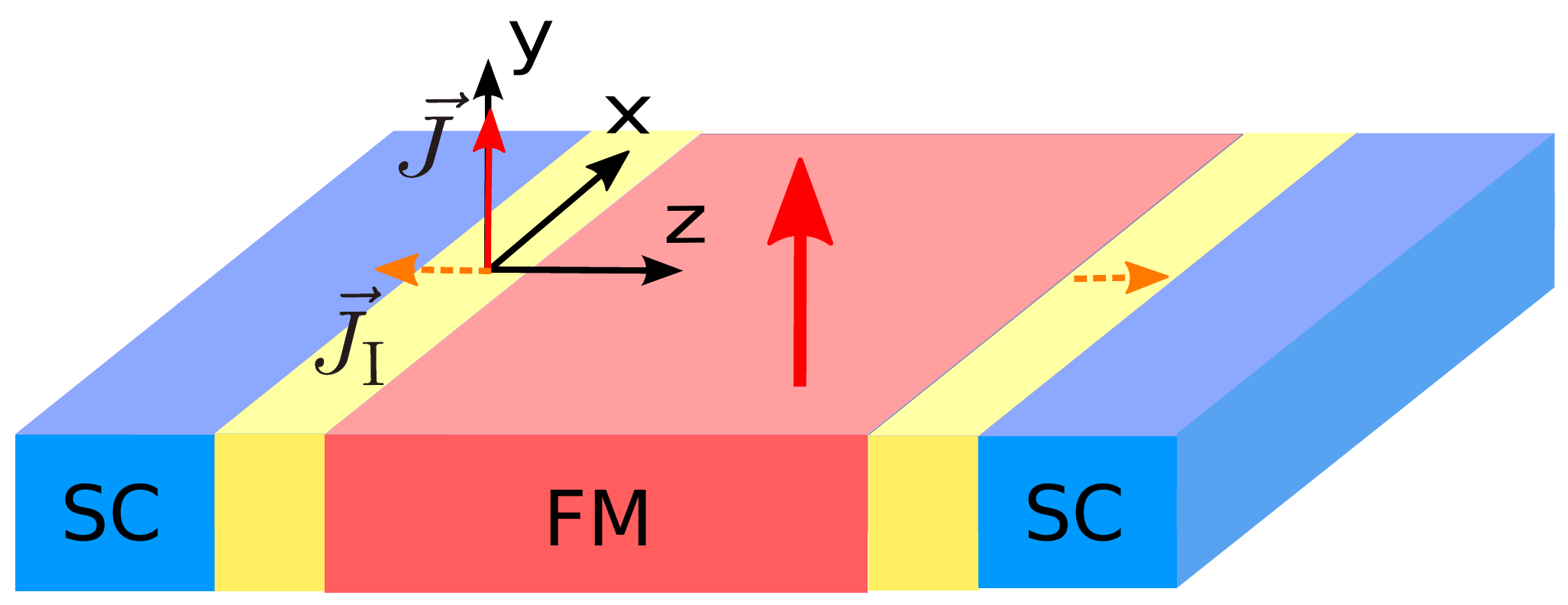}
\caption{\label{sketch} The setup studied here is a ferromagnetic Josephson junction. Blue regions indicate the superconducting electrodes, red regions the central ferromagnetic layer. Yellow regions are spin-active interfaces which feature a magnetization misaligned with that of the FM bulk.}
\end{figure}

\section{Quasiclassical Theory of Superconductivity}

Let us first summarize the theory for a system where the band structure is not spin-polarized.
The quasiclassical Green function is obtained from Gor'kov's Green function\cite{gorkov58} by an integration over momentum\cite{eilen, larkin68, Serene, rammer86}:
\begin{equation} 
\label{xiintegration}
\hat{g}(\vec{p}_{\rm{F}} , \vec{R}, \epsilon_n, t)=\frac{1}{a(\vec{p}_{\rm F})}\int {\rm d}\xi_p \hat{\tau}_3 \hat{G}(\vec{p}, \vec{R}, \epsilon_n, t),\end{equation}
where $a(\vec{p}_{\rm F})$ is a quasiparticle renormalization parameter, and $\xi_p=\vec{v}_{\rm{F}} \cdot (\vec{p}-\vec{p}_{\rm{F}} )$. The ``hat'' denotes a 4$\times$4 matrix in combined spin- and Nambu-Gor'kov space. The propagator $\hat{g}$ depends on the spatial coordinate $\vec{R}$, time $t$, Matsubara-frequency\cite{matsubara55} $\epsilon_n=2\pi T(n+\frac{1}{2})$ and the momentum $\vec{p}_\mathrm{F}$, which lies on the Fermi-surface. The Fermi velocity corresponding to Fermi momentum $\vec{p}_{\rm F}$ is denoted by $\vec{v}_{\rm F}$. The resulting equation of motion, which can be obtained from the Dyson-equation of Gor'kov's Green function, is the Eilenberger-equation\cite{eilen}:
\begin{equation} \label{eilen1} i \hbar \vec{v}_{\rm{F}} \cdot \nabla_{\vec{R}}\hat{g}+[i\epsilon_n\hat{\tau}_3-\hat{\Delta}-\hat{\Sigma}, \hat{g}]=\hat{0},\end{equation}
which must be supplemented by a normalization condition:
$\hat{g}\circ \hat{g}=-\hat{1}\pi^2$. $\hat{\Delta}$ denotes off-diagonal self-energies in particle-hole space and $\hat{\Sigma}$ the diagonal ones. 

The equilibrium current density in a system without spin polarization reads:
\begin{equation}\vec{j}(\vec{R},t)=e N_{\rm F}k_\mathrm{B} T \sum\limits_{\epsilon_n}\frac{1}{2}
\mathrm{Tr}_4
\Big\langle \vec{v}_{\rm{F}}(\vec{p}_{\rm F})
\hat{\tau}_3\hat{g} (\vec{p}_{\rm{F}},\vec{R}, \epsilon_n, t)\Big\rangle, \end{equation}
where Tr$_4$ denotes a trace over spin and particle-hole (Nambu-Gor'kov) degrees of freedom, and $\langle\bullet \rangle$ and $N_\mathrm{F}$ are defined by:
\begin{eqnarray}
\langle \bullet\rangle&=&\frac{1}{N_{\mathrm{F}}}\int_{FS} \frac{{\rm d}^2 p_{\mathrm{F}}}{(2\pi\hbar)^3|\vec{v}_{\mathrm{F}}(\vec{p}_{\mathrm{F}})|} \; (\bullet )\; ,\\
N_{\rm F}&=&\int_{FS} \frac{{\rm d}^2 p_{\mathrm{F}}}{(2\pi\hbar)^3|\vec{v}_{\mathrm{F}}(\vec{p}_{\mathrm{F}})|}.
\end{eqnarray}
$N_{\rm F}$ is the density of states per spin at the Fermi level. In appendix \ref{FSA} we detail our numerical Fermi-surface averaging procedure.

For a ferromagnet these expressions need to be modified in order to take into account the different Fermi surfaces for the two spin channels.
As discussed in our earlier work\cite{grein09,grein10}, we model the ferromagnet by two spin-scalar quasiclassical Green functions for each band. This is consistent with the quasiclassical approximation in the limit $J>0.1\ E_\mathrm{F}$ and naturally results in a three-channel matching problem at the SC/FM interface. We underline that an equivalent approach in the diffusive limit of quasiclassical theory (Usadel-equation \cite{usadel70}) is currently not available for lack of multi-channel boundary conditions. All definitions given above remain the same apart from the reduced matrix structure of the Green function and the fact that $\vec{v}_\mathrm{F\eta}$, $\vec{p}_\mathrm{F\eta}$, $N_\mathrm{F\eta}$ and $\langle\bullet\rangle_\eta$ now depend on the spin band $\eta$ in question.

We will make extensively use of 
a fundamental symmetry relating particle-like and hole-like quantities in quasiclassical theory, for which we introduce the $\tilde\, $-operation defined by
\begin{align} \tilde Q(\vec{p}_\mathrm{F}, \vec{R}, \epsilon_n,t)=[Q(-\vec{p}_\mathrm{F}, \vec{R}, \epsilon_n,t)]^*.\end{align}

\subsection{Riccati parameterization}
We use a particular Riccati parameterization of the quasiclassical (QC) Green function\cite{schopohl95,EschrigThesis,eschrig99,eschrig00, eschrig09} which has proven very useful in the past. For Matsubara's Green functions,\cite{matsubara55} it consists of 2 parameters (``coherence functions'') $\gamma$, $\tilde\gamma$, describing particle-hole coherence in the superconducting state. The Green function $\hat g$ reads in terms of these parameters\cite{eschrig09}:
\newcommand{\plus}{\;\;\,}
\newcommand{\mat}{\left( \begin{array}{cc} }
\newcommand{\matend}{\end{array}\right)}
\begin{equation}
\label{cgretav}
\hat g=
\mp \, 2\pi i\,
\mat \plus {\cal G} & \plus {\cal F} \\ -\tilde{\cal F} &
-\tilde{\cal G} \matend
\pm i\pi \hat \tau_3
,
\end{equation}
where ${\cal G}=({\it 1}-\gamma \tilde\gamma )^{-1}$, ${\cal F}={\cal G} \gamma $, $\mp$ and $\pm$ correspond to $\epsilon_n>0$ and $\epsilon_n<0$, respectively. The transport equations are:
\begin{align} (i\hbar \vec{v}_{\rm F}\cdot \nabla+2i\epsilon_n)\gamma=[\gamma \tilde \Delta \gamma+\Sigma \gamma-\gamma \tilde \Sigma-\Delta],\\
(i\hbar \vec{v}_{\rm F}\cdot \nabla-2i\epsilon_n)\tilde \gamma=[\tilde \gamma \Delta \tilde \gamma+\tilde \Sigma \tilde \gamma-\tilde \gamma \Sigma-\tilde \Delta].\nonumber\end{align}
For a superconductor or normal metal
the $\gamma $, $\tilde \gamma $, $\Delta $, $\tilde \Delta $, $\Sigma$, and $\tilde \Sigma $ are 2x2 spin matrices for each Fermi surface point $\vec{p}_{\rm F}$, as all Fermi surfaces can be regarded as doubly spin degenerate on the scale of the Fermi energy. 
A weak band splitting due to an external field or as in ferromagnetic alloys can be incorporated as source field in these equations, leading to spin-dependent self energies.
In contrast, for a ferromagnet 
with strong spin splitting of the energy bands the transport equations hold separately for each spin band $\eta $ with corresponding
Fermi surface point $\vec{p}_{{\rm F}\eta }$, and the 
above-mentioned quantities are all scalars. The important difference lies in the integration over $\xi_p $ in equation \eqref{xiintegration}, which destroys all coherence between quasiparticle excitations living on different spin bands for the latter case of strong spin polarization, however allows for quasiparticle coherence between spin bands when the band splitting is on the low-energy scale that constitutes the phase space for quasiparticles.

The self-energies introduced here are related to those we used in Eq.~\eqref{eilen1} by
\begin{align} [\hat{\Delta}+\hat{\Sigma}]=\left(\begin{array}{cc} \Sigma & \Delta \\  \tilde \Delta & \tilde \Sigma \end{array}\right). \end{align}

There is a further symmetry relating  the Matsubara coherence functions with positive and negative frequencies:
\begin{align} \gamma(-\epsilon_n)=[\tilde \gamma(\epsilon_n)]^\dagger. \end{align}
As a result, we only need to consider  $\gamma(\epsilon_n>0)$ in the following. All other quantities can then be obtained from the symmetry relations. The current density in the FM spin-bands ($\eta\in 2,3$) is obtained from
\begin{align}\vec{j}_{\eta}(\vec{R},t)=4e N_{\rm F\eta}k_\mathrm{B} T \sum\limits_{\epsilon_n>0}
\Big\langle \vec{v}_{\rm{F}\eta}(\vec{p}_{\rm F\eta}) {\rm Re}\big[g_\eta\big]\Big\rangle_{\eta+},\end{align}
with $g$ the $11$-component of $\hat{g}_\eta$ and $\langle \bullet \rangle_{\eta+}$ denotes a Fermi-surface average which is taken over momenta with positive projection on the $z$-axis only. To obtain this expression, we did not only exploit the symmetry relations but also the fact that for the spin channels an additional symmetry between the diagonal components of $\hat{g}$ holds: $g_\eta(\epsilon_n,p_{\rm F\eta})=-\tilde g_\eta(\epsilon_n,p_{\rm F\eta})$.

\subsection{Boundary conditions}

The boundary conditions for Riccati amplitudes at interfaces and surfaces are formulated in terms of the normal state scattering matrix $S$. \cite{eschrig00, eschrig09} The transport equations for Riccati amplitudes can be solved by integrating them along their characteristics, which are straight lines in real space\cite{eschrig00}. Every amplitude has a unique stable direction of integration. One may hence group them into ``incoming'' and ``outgoing'' amplitudes with respect to a scattering region, which will be the interfaces here. The convention is that outgoing amplitudes are denoted by capital case letters and incoming ones by small case letters. The transport channels that participate in scattering are labeled by $k,k'$. In our case, these will be band indices, as the conservation of parallel momentum that we assume excludes the scattering between different momenta in the same band. The boundary conditions then relate outgoing to incoming amplitudes. 
For numerical calculations, the most convenient way of solving these boundary conditions is to calculate\cite{eschrig09}:
\begin{align}
\boldsymbol{\mathcal{F}}=(1-\tilde{\boldsymbol{\gamma}}\circ\boldsymbol{\gamma}')^{-1}\circ\boldsymbol{\gamma}',\quad \boldsymbol{\mathcal{G}}=(1-\boldsymbol{\gamma}'\circ\tilde{\boldsymbol{\gamma}})^{-1},\end{align}
where $\circ$ denotes a matrix product in channel space and:
\begin{align}\tilde{\boldsymbol{\gamma}}_{kk'}=\delta_{kk'}\tilde\gamma_k, \quad \boldsymbol{\gamma}_{kk'}=\delta_{kk'}\gamma_k,\quad\boldsymbol{\gamma}'=S\circ\boldsymbol{\gamma}\circ\tilde S.\end{align}
$S$ is the normal state scattering matrix, which -- thanks to the conservation of parallel momentum -- is block-diagonal here, and the blocks can be labeled by $k_{||}$. The outgoing Riccati amplitudes are then obtained from:
\begin{align} \Gamma_k=\boldsymbol{\mathcal{G}}^{-1}_{kk}\boldsymbol{\mathcal{F}}_{kk}. \end{align}
As explained at length in our earlier publications,\cite{eschrig09,grein10} the $S$-matrices will be either $4 \times 4$ or $3 \times 3$, depending on whether $k_{||}$ is larger or smaller than the Fermi wave vector of the minority band in the FM. Thus, all matrices defined above will have these dimensions, as channels with different $k_{||}$ do not mix. 

\section{Solution of the Eilenberger equation}

\subsection{Impurities}
The simplest way to consider impurity scattering is to look at elastic, non-magnetic scattering in the self-consistent Born-approximation. This gives the following self-energies in the ferromagnet\cite{eschrig08}:
\begin{align} \left(\begin{array}{cc} \Sigma & \Delta \\ \tilde \Delta & \tilde \Sigma\end{array}\right)_\eta=\frac{1}{2\pi \tau_\eta}\langle\hat g_\eta\rangle_\eta. \end{align}
Note that here $\Delta$, $\tilde \Delta$ denote all off-diagonal self-energies in Nambu-space, rather than a superconducting gap. The Riccati transport equations then need to be solved numerically, and in each step the impurity self-energy $1/(2\pi\tau_\eta)\langle\hat g_\eta\rangle_\eta$ is updated.

The scattering time $\tau$ is related to the scattering length $l$ by $l=v_{\rm F}\tau$. Since $l$ corresponds to the mean free path of the quasiparticles, it is directly related to the impurity concentration and should be the same for both spin-bands (we consider here spin-inactive scattering). The Fermi-velocities, however, are different and hence $\tau_2=l/v_{\rm F,2}<\tau_3=l/v_{\rm F,3}$.
This argument can also be made quantitative by noting that the impurity self energy in Born approximation is proportional to $c_iV_i^2N_{\rm F}$, with impurity concentration $c_i$, impurity scattering potential $V_i$, and density of states at the Fermi level $N_{\rm F}$. As this defines $1/\tau$, one obtains for the two spin directions $\tau_\eta \sim 1/N_{{\rm F},\eta}$. As shown in appendix \ref{FSA}, $N_{{\rm F},\eta}\sim m^2v_{{\rm F},\eta}$, and thus (assuming equal effective masses for the spin bands) $v_{{\rm F},2}\tau_2=v_{{\rm F},3}\tau_3$, i.e. the mean free path $l$ for both spin bands is equal.

\subsection{Gap-equation}
The mean-field gap-equation in the SC is
\begin{align} \Delta(z)\cdot\ln\left(\frac{T}{T_c}\right)=\sum_n\left( \langle f(\epsilon_n,\vec{k}_{\rm F},z)\rangle-\pi\frac{\Delta}{|\epsilon_n|}\right),\end{align}
where the coupling constant is eliminated in favor of $T_c$. 
In appendix \ref{NUM} we discuss in detail some sophisticated problems when iterating this equation at an interface.
We show that whenever the quasiclassical Green function undergoes a jump-discontinuity under reflection from the interface, 
the only self-consistent solution of the above equation at the interface is $\Delta=0$. The discontinuity of the quasiclassical propagator in turn results from the fact that the microscopic reflection process at the interface falls outside the range of applicability of QC theory. To remedy this issue, we introduce a length cut-off $\xi_c\ll \xi_0$ and calculate the gap-equation only for distances larger than $\xi_c$ from the interface. We show that if the cut-off length is chosen small enough, neither its precise value nor the profile of the SC-gap for distances smaller than $\xi_c$ influences the results. We choose for our numerical calculations $\xi_c=0.01\ \xi_0$.

\subsection{Numerical solution}
We found that a very stable way of solving the Riccati differential equations for spin-polarized systems is to resort to analytical solutions for constant self-energies. I.e., instead of using a standard solver for ordinary differential equations along each trajectory, we approximate the self-energies by a constant in each interval that corresponds to the grid-spacing of our discretization in the $z$-direction, and use the analytical solution for homogeneous self-energies \cite{eschrig09}:
\begin{align}\gamma(\rho)=\gamma_h+e^{i\rho\Omega_1}[\gamma_0-\gamma_h]\Big\{e^{i\rho\Omega_2}+C(\rho)\cdot[\gamma_0-\gamma_h]\Big\}^{-1},\end{align}
where $\gamma_h$ is the homogeneous bulk solution,
$\gamma_0$ is the initial value and $\rho$ parameterizes the respective trajectory as $\vec{R}=\vec{v}_{\rm F}(\vec{k}_{\rm F})\rho+\vec{R}_0$. 
Moreover, $\Omega_1=i\epsilon_n-\Sigma-\gamma_h\tilde\Delta$, $\Omega_2=-i\epsilon_n-\tilde\Sigma+\tilde \Delta \gamma_h$ and 
\begin{align} C(\rho)=C_0 e^{i\rho\Omega_1}-e^{i\rho\Omega_2}C_0,\end{align}
where $C_0$ is given by the solution of:
\begin{align} \tilde \Delta=C_0\Omega_1-\Omega_2 C_0.\end{align}
The value of $\gamma$ at grid point $n$ is then calculated from the analytical solution with the value at grid point $n-1$ as initial value.
These general formulas simplify inside the FM, because all quantities are scalar and commute. In the SC they simplify as well, since we only consider the order parameter self-energy there and thus always have $\Delta,\ \tilde\Delta,\ \gamma_h\propto i\sigma_y$ and $\Sigma,\ \tilde\Sigma=0$.
The homogeneous bulk solution is for these cases
\begin{align} \gamma_h=-\frac{\Delta}{\mathcal{E}\pm i\sqrt{-\Delta\tilde\Delta-\mathcal{E}^2}}, \end{align}
where $\mathcal{E}=i\epsilon_n-(\Sigma-\tilde \Sigma)/2$. 

\begin{figure}[t]
\includegraphics[width=6cm]{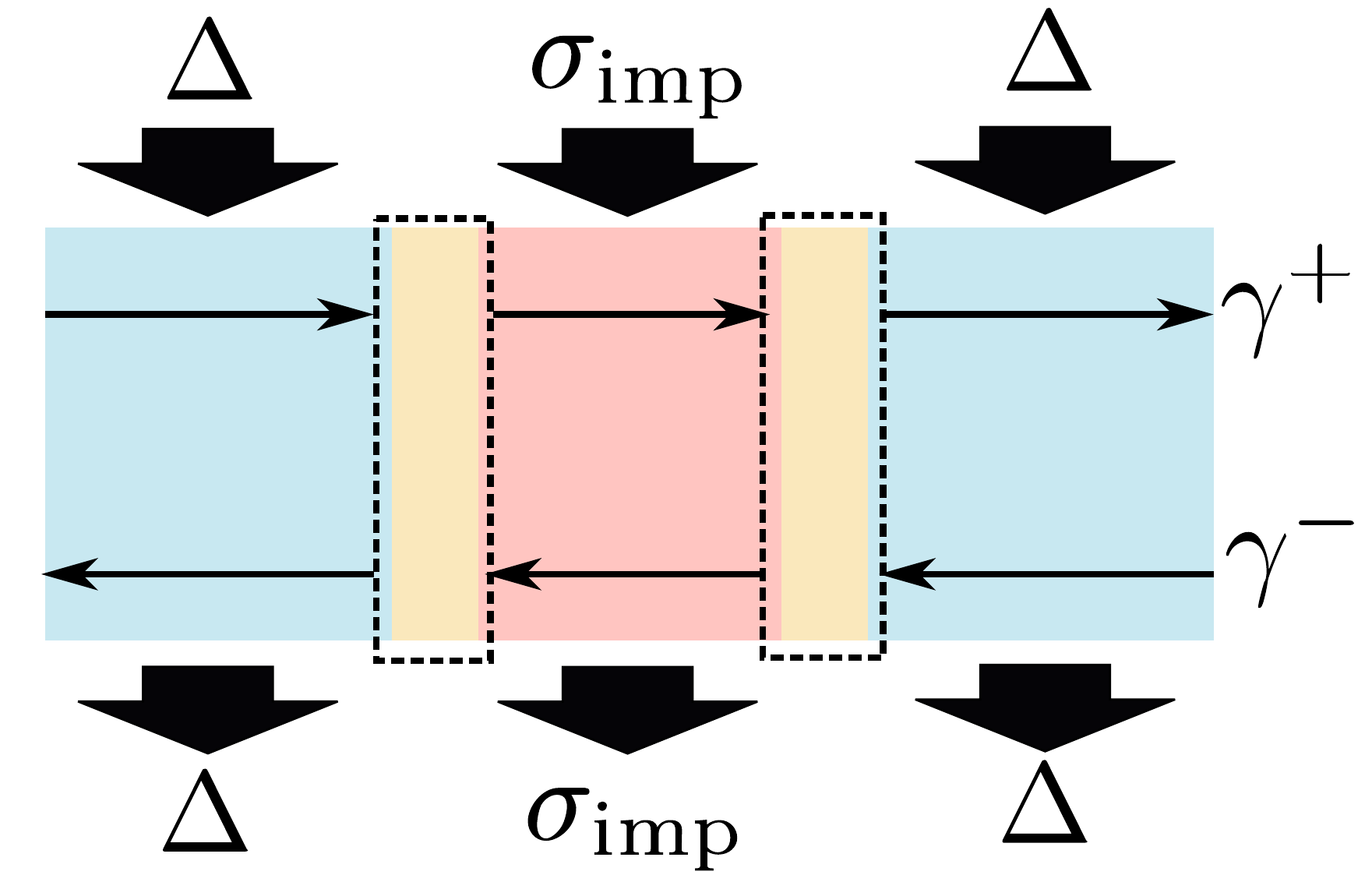}
\caption{\label{setup} Illustration of the iterative procedure for obtaining a fully self-consistent solution.}
\end{figure}

As we include
the self-consistent equations for the impurity potential and the order parameters, we have to deal with a non-linear problem, in which the solutions for different trajectories and Matsubara frequencies are coupled. We solve this problem iteratively as 
as illustrated in Fig.~\ref{setup} (see also appendix \ref{IS}).
At each step, we assume a given set of initial values for the coherence functions at the outer boundaries of the system and at the interfaces and given values for the impurity potential and order parameters. We then integrate the transport equations and solve the boundary conditions at the interfaces to update all coherence functions and calculate the initial values for the next step of the iteration. Subsequently, all self-energies are updated by evaluating the respective self-consistent equation with the new coherence functions. The initial values at the outer boundaries are always assumed to be the respective SC bulk-solution. The order-parameter phase difference $\Delta\chi$ enters via these boundary values. At the initial step, we chose the coherence functions and the impurity potential inside the FM to be zero and the SC gap to be constant up to the interface.

\section{Scattering Matrix}

The normal-state scattering matrices entering the quasiclassical boundary conditions cannot be obtained within QC theory itself. We calculate them from plane wave matching in the normal-state using a simple model for the interface scattering potential that captures all relevant effects.

The interface Hamiltonian reads:
\begin{align} 
H^{\rm I}=\sum\limits_{\vec{k}\mu\nu}
c^\dagger_{\vec{k}\mu}(\xi_{\vec{k}}^{\rm I}
\delta_{\mu\nu}+\vec{h}\cdot \vec{\sigma })c_{\vec{k}\nu},
\end{align}
where $\vec{h}$ is the interface exchange field, and:
\begin{align} \xi_{\vec{k}}^{\rm I}=\frac{\hbar^2 \vec{k}^2}{2m}+V_I-E_{\rm F}.\end{align}
$V_I$ plays the role of a spin-independent interface potential.
Accordingly, the normal-state Hamiltonians of the ferromagnet and the superconductor read:
\begin{align}  H^{\rm FM}&=\sum\limits_{\vec{k}\mu\nu}c^\dagger_{\vec{k}\mu}(\xi_{\vec{k}}^{\rm FM}\delta_{\mu\nu}+J\cdot \vec{\sigma})c_{\vec{k}\nu},\\
H^{\rm SC}&=\sum\limits_{\vec{k}\mu\nu}c^\dagger_{\vec{k}\mu}(\xi_{\vec{k}}^{\rm SC}\delta_{\mu\nu})c_{\vec{k}\nu}.\end{align}
 Obviously, $H^{\rm SC}$ is invariant under rotations in spin-space, while both $H^{\rm I}$ and $H^{\rm FM}$ break spin-rotation symmetry. For $\xi_{\vec{k}}^{\rm I, SC, FM}$ we assume for simplicity a parabolic dispersion with the same effective mass in all cases. The dispersions $\xi^{{\rm I}}_{\vec{k}}$ and and $\xi^{\rm FM}_{\vec{k}}$ are shifted by a constant energy $V^{\rm I}$ and $V^{\rm FM}$, respectively, compared to $\xi^{\rm SC}_{\vec{k}}$.
The unitary matrix $U$, which maps the spin-eigenbasis of the interface to that of the ferromagnet, reads:
\begin{align} U=\left(\begin{array}{cc} \cos\frac{\alpha}{2} & -\sin \frac{\alpha}{2}e^{-i\vphi}\\ \sin \frac{\alpha}{2} e^{i\vphi} & \cos \frac{\alpha}{2}\end{array}\right). \end{align}

The scattering matrix is obtained by diagonalizing the Hamiltonians and inferring the respective values of $k_z$ at the Fermi-level for a given $k_{||}$. The corresponding wave-functions in the three layers are then matched at the SC/I and the I/FM interface respectively, choosing a spin-quantization axis in the SC aligned with that of the interface, while at the I/FM interface, the rotation $U_{\vec{k}}$ must be taken into account.

The requirement that $k_{||}$ must be conserved across the interface implies that even if the ferromagnet is not fully spin-polarized, some trajectories will have an evanescent solution in the minority band of the FM, corresponding to the half-metallic case. In the case were this minority band solution is propagating we obtain a $4\times4$-scattering matrix, if it is evanescent, the matrix is $3\times3$ as discussed in appendix \ref{SM}.

\section{Results}
\begin{figure}[t]
\includegraphics[width=\columnwidth]{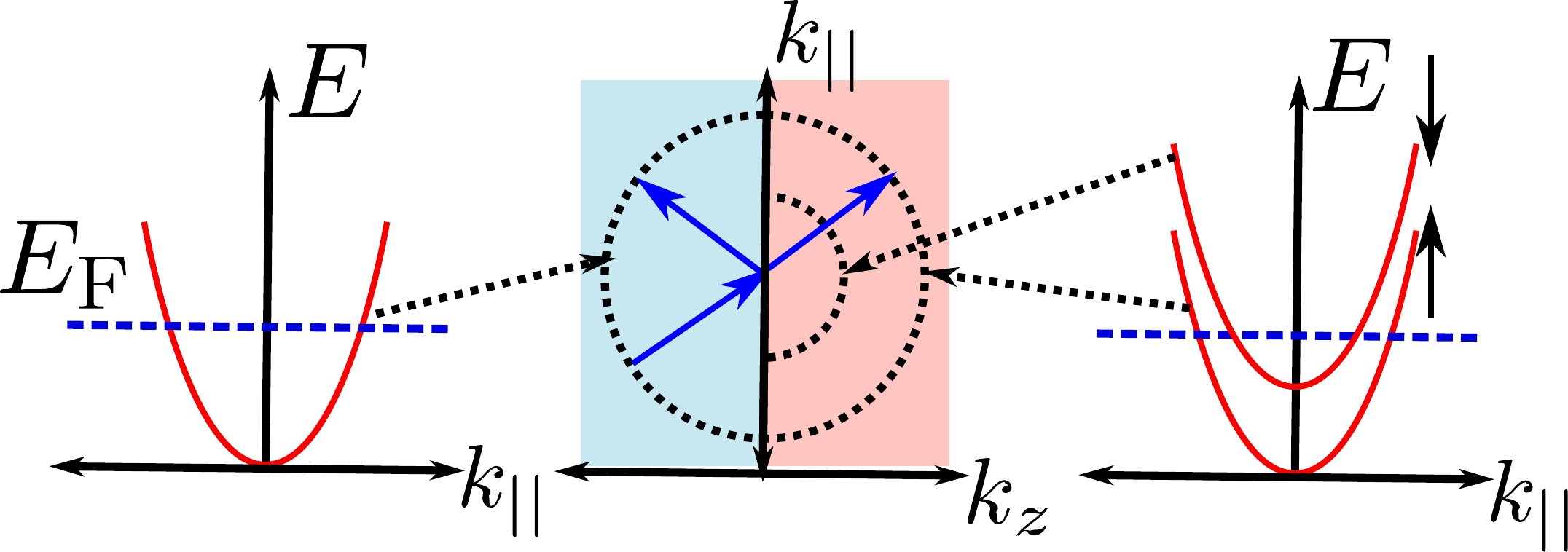}
\caption{\label{scattering} Scattering geometry considered here. The majority band of the FM is assumed to have the same Fermi-momentum as the superconductor.}
\end{figure}
Our scattering geometry is shown in in Fig.~\ref{scattering}. In what follows, the Fermi-momentum of the majority (spin-up) band is assumed to be identical to the Fermi-momentum of the SC. This is simply a matter of convenience, as we must only distinguish two and not three types of trajectories in this case. 
The third type of trajectories would describe total reflection from the interface, which does not contribute to transport. Our methods can easily applied to the general case as well.
The value of the minority band Fermi-momentum is determined by the exchange field $J$, the magnitude of which is assumed to be the same in the interface and in the FM bulk (however not the direction).

\begin{figure}[t]
\includegraphics[width=\columnwidth]{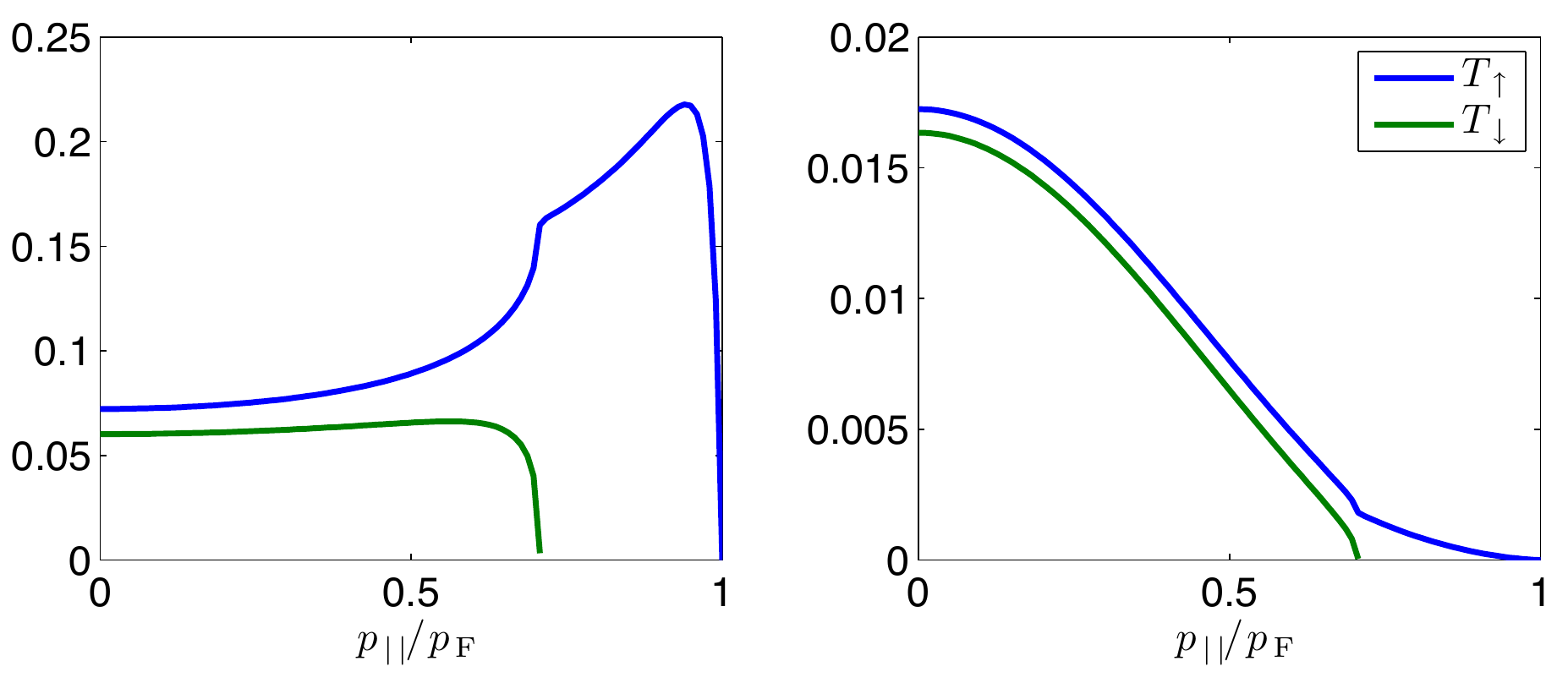}
\caption{\label{fig0} The triplet transmission parameters $T_{\uparrow,\downarrow}$ as a function of $p_{||}$, for the high-transparency interface ($d=0.25\ \lambda_{\rm F}/2\pi$, $V_{\rm I}=1\ E_{\rm F}$) (left), and the tunneling interface ($d=2\ \lambda_{\rm F}/2\pi$, $V_{\rm I}=2\ E_{\rm F}$) (right). Other parameters are $J=0.5\  E_{\rm F}$, $\alpha=0.5\ \pi$.}
\end{figure}

In what follows we consider two parameter sets for the interface. A ``high''-transparency interface with thickness $d=0.25\ \lambda_{\rm F}/2\pi$ and $V_I=1\ E_{\rm F}$ and a ``low''-transparency interface with $d=2\ \lambda_{\rm F}/2\pi$, and $V_I=2\ E_{\rm F}$. If $S_{\uparrow,{\rm SC}}=(t_{\uparrow\uparrow}\; ,\; t_{\uparrow\downarrow})$ and $S_{{\rm SC},\uparrow}=(t'_{\uparrow,\uparrow}\; ,\; t'_{\uparrow\downarrow})^T$ are the sub-matrices of $S$ related to transmission from the SC to the spin-up band of the FM and vice versa, then the relevant transmission quantity for inducing triplet correlations in the spin-up band is:
\begin{align} T_\uparrow=|S_{\uparrow,{\rm SC}} i\sigma_y S_{{\rm SC},\uparrow}|, \end{align}
and analogously we have $T_\downarrow$ for the spin-down band. These quantities are plotted in Fig.~\ref{fig0} for the two interfaces discussed here. Their functional dependence on $k_{||}$ was investigated extensively in Ref.~\onlinecite{grein10}.

In order to discuss various symmetry components of the induced triplet pair correlations, We use
the following decomposition of the quasiclassical Green functions: 
\begin{align} g=g_0+\vec{\sigma}\cdot\vec{g},\quad f=(f_s+\vec{\sigma}\cdot \vec{f_t})i\sigma_y,\end{align}
and obtain
the induced magnetization from
\begin{align} \vec{M}=2\mu_{\rm B}k_{\rm B} T N_{\rm F}\sum_n \langle \vec{g}(\epsilon_n,\vec{p}_{\rm F})\rangle.\end{align}
As measures of odd-frequency and odd-momentum triplet amplitudes we define
\begin{align} \vec{f}_\epsilon&=\frac{k_{\rm B} T}{2}\sum_{n>0} \langle \vec{f}_t (\epsilon_n,\vec{p}_{\rm F})-\vec{f}_t(-\epsilon_n,\vec{p}_{\rm F})\rangle,\\
\vec{f}_p&=\frac{k_{\rm B} T}{2}\sum_{n}\Big[ \langle \vec{f}_t (\epsilon_n,\vec{p}_{\rm F})\rangle_+-\langle\vec{f}_t(\epsilon_n,\vec{p}_{\rm F})\rangle_-\Big].\end{align}
Here, $\vec{f}_\epsilon$ is the s-wave component of the odd-frequency correlations. $\vec{f}_p$ does not correspond to a particular spherical harmonic. Our model system only breaks translational invariance in the $z$-direction and is rotationally invariant in the $x$-$y$-plane. We therefore project out the correlations which are odd under $p_z\mapsto -p_z$.

\begin{figure}[t]
\includegraphics[width=\columnwidth]{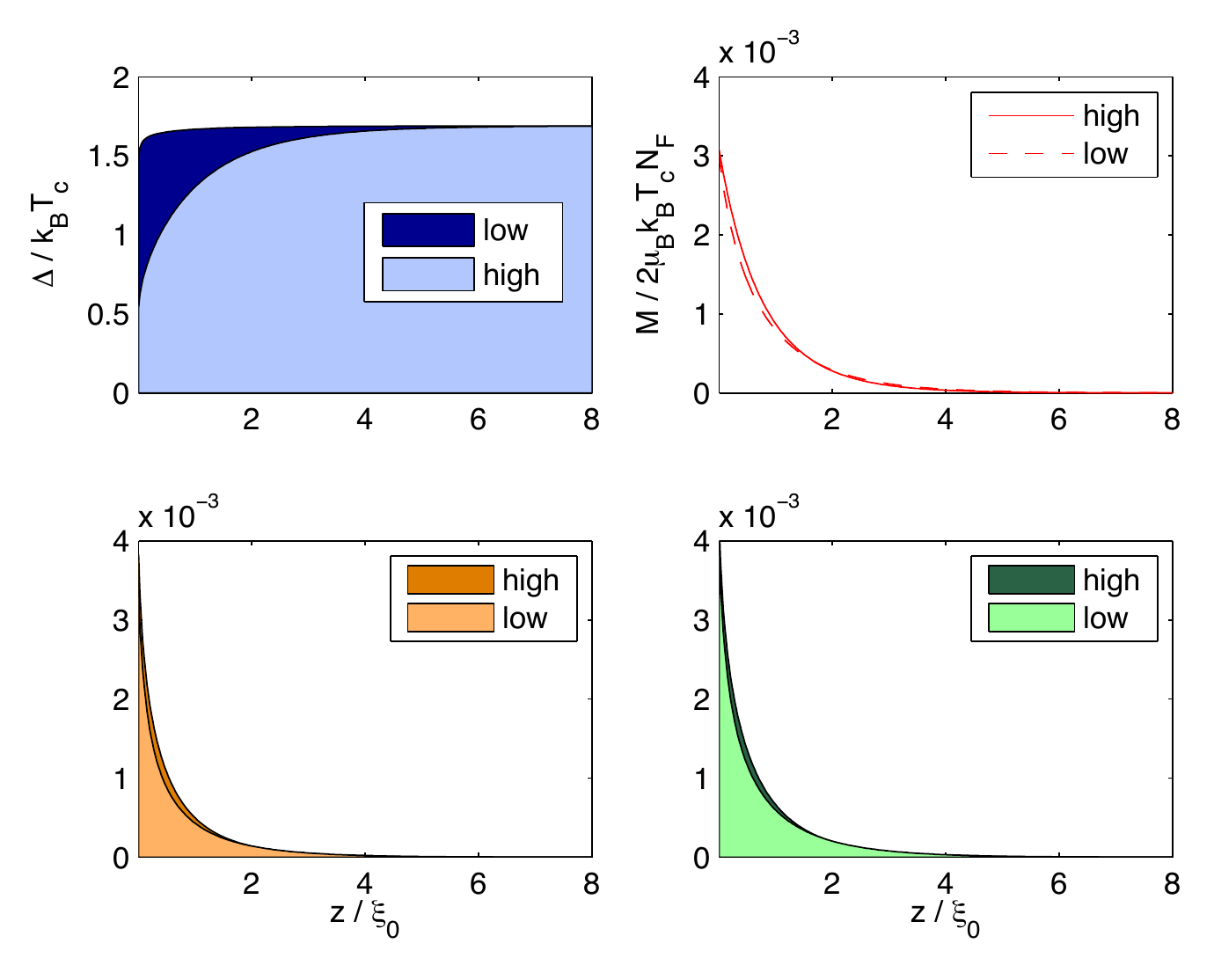}
\caption{\label{inverse} Inverse proximity effect in the SC-leads -- ``high'' refers to the high transparency interface and ``low'' to the low transparency interface.  The top row plots show the spatial profile of the SC-gap $\Delta $ (left) and the induced magnetization $M=|\vec{M}|$ (right). At the bottom row, the induced triplet correlations which are odd in momentum (right, green) or odd in frequency (left, orange) are shown. Parameters are $l=100 \xi_0$, $L=0.5 \xi_0$, $\alpha=0.5\pi$, $T=0.5\ T_c$, $\Delta\chi=0$, $z\in[\xi_c,8 \xi_0]$.} 
\end{figure}
\begin{figure}[t]
\includegraphics[width=\columnwidth]{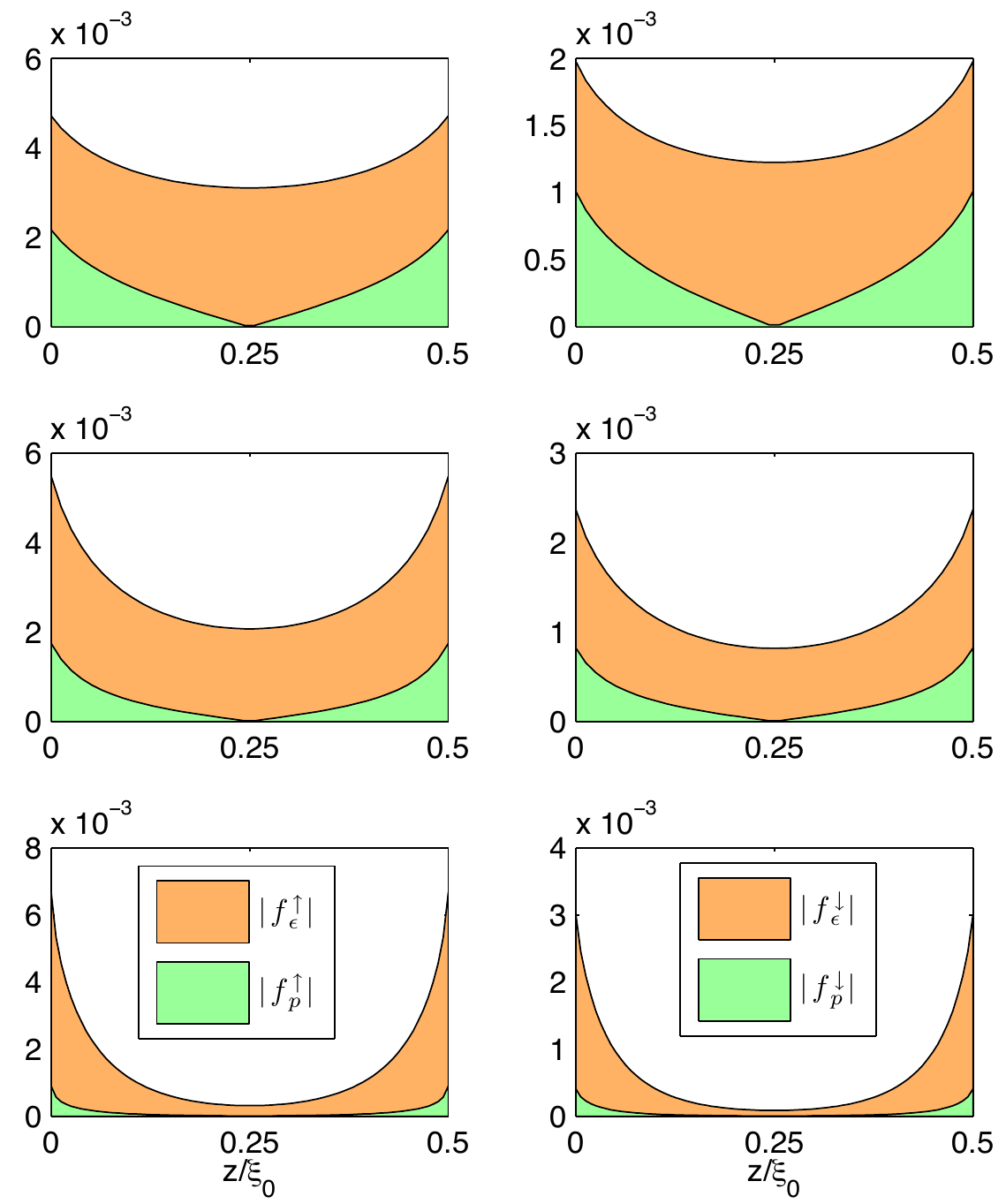}
\caption{\label{correlations} Spatial profile of equal-spin triplet-correlations -- The plots show the total momentum averaged triplet correlations $f^\eta_\epsilon $ (s=wave, odd frequency) and the correlations odd in momentum (even frequency) $f^\eta_p$, with $\eta=\uparrow,\downarrow$, inside the FM-layer at the lowest Matsubara frequency.  The interfaces are located at $z=0$ and $z=0.5\ \xi_0$. $d=0.25\ \lambda_{\rm F}/2\pi$, $T=0.5\ T_c$, . 
The mean free path for the three rows are from top to bottom: $l=\xi_0$, $l=0.1\xi_0$, and $l=0.01\xi_0$.
Other parameters as in Fig.~\ref{inverse}.}
\end{figure}
\begin{figure}[b]
\includegraphics[width=\columnwidth]{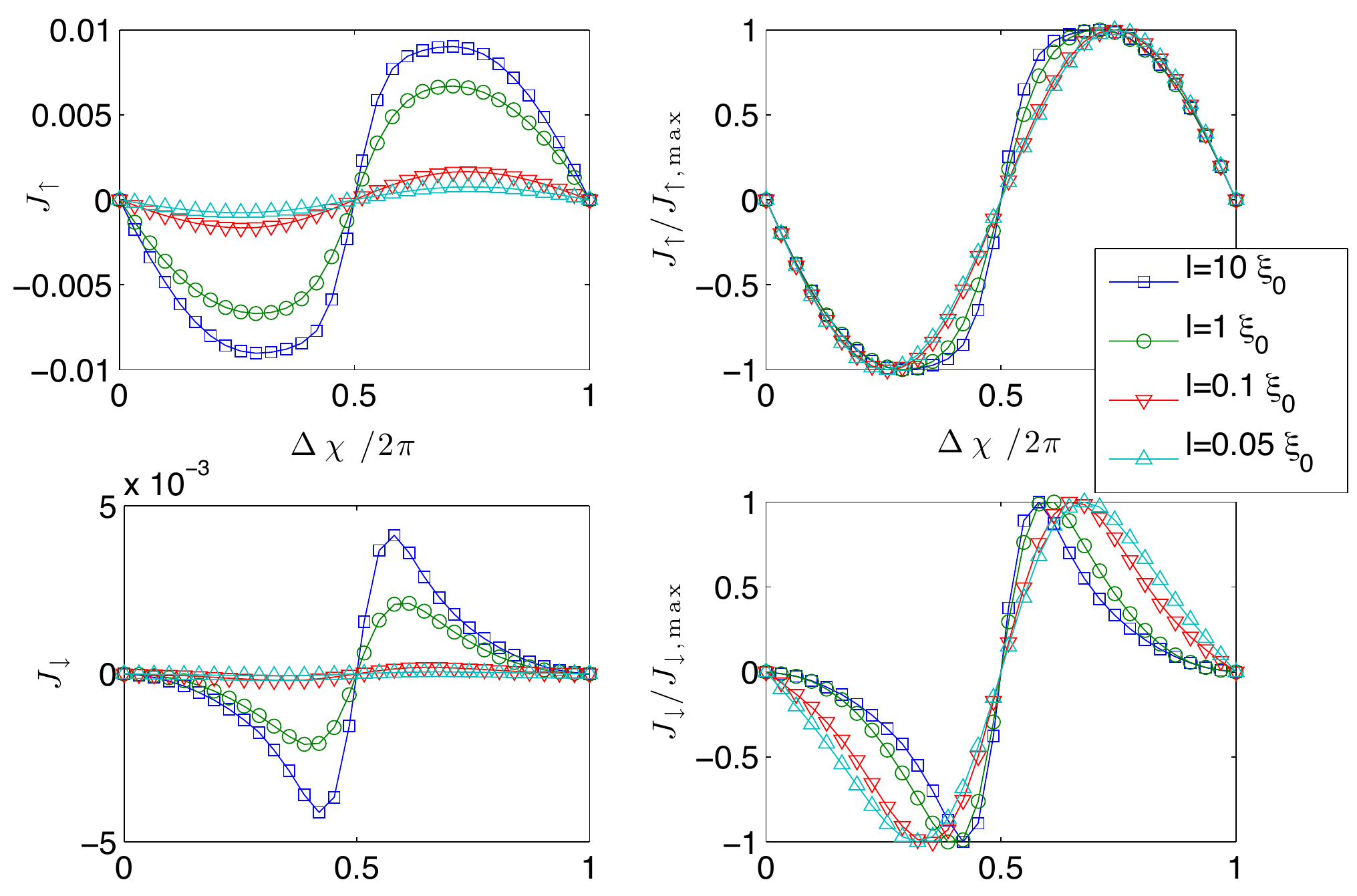}
\caption{\label{impurities} Influence of disorder -- current-phase relation of spin-up (top) and spin-down (bottom) currents for different scattering lengths. Data is shown unnormalized (left column) and normalized to its maximum value (right column). Other parameters as in Fig.~\ref{inverse} for high transparency. The current density is stated in units of $e k_{\rm B} \Tc (\pF/2\pi)^2/\hbar^3$.}
\end{figure}
We first investigate the inverse proximity effect, as shown in Fig.~\ref{inverse}. Spin-active scattering at the SC/FM interface induces triplet-pairing correlations and a magnetization inside the SC-electrodes close to the interface. Moreover, the order parameter is suppressed at the interface. While this is also caused by spin-active scattering, here the predominant cause for this suppression is the transparency of the interface. This can be seen in the topmost plots of Fig.~\ref{inverse} (left), where we compare a highly transparent interface to an interface in the tunneling limit. The reason why the spin-mixing effect plays a minor role here, is that under the assumption of a box shaped scattering potential the magnitude of the interface spin-mixing parameters that are responsible for creating triplet-pair correlations are typically small (see our extensive discussion of this effect in Ref.~\onlinecite{grein10}). This can be seen in the top right panel, where the absolute value of the induced magnetization is shown. The transparency is largely irrelevant here, as the inverse proximity effect is due to spin-active back-scattering from the superconducting side, i.e. a property of the spin-dependent scattering phases in the reflection parts of the scattering matrix. 
The induced magnetization decays on the scale of the superconducting coherence length, $\xi_0$. Note that this induced magnetization is a result of the superconducting inverse proximity effect (production of triplet pair correlations by the magnetic interface) and absent in the normal state.

In the lower panels of Fig.~\ref{inverse} we project our various symmetry components of the triplet pair correlations. 
Since the SC itself is spin-rotation invariant, it makes sense to plot the invariant quantity $f_t=|\vec{f}_t|$ rather than the individual components of $\vec{f}_t$, which are not invariant. Since we consider in Fig.~\ref{inverse} a rather clean superconductor ($l=100\xi_0$), odd-momentum and odd-frequency correlations are comparable in magnitude and decay on the same length scale $\xi_0$. 
In the case of the high transparency interface, the induced triplet-correlations and the magnetization show a slight dependence on the length of the junction (not shown), i.e. superconducting correlations that propagated through the FM-layer also contribute to the inverse proximity effect in that case.

We now study the influence of impurity scattering on the triplet pair correlations.
In Fig.~\ref{correlations}, we plot both the spatial profiles of the absolute value of the total momentum averaged equal-spin triplet correlations, $|f^\eta_\epsilon |$ (s-wave, odd frequency) and of the correlations which are odd under $p_z\mapsto -p_z$, $|f^\eta_p|$ (odd momentum), for each spin direction inside the FM-slab. 
As the impurity concentration increases, these correlations decay faster into the FM and the relative magnitude of $|f^\eta_p|$ decreases. However, the magnitude of the latter only becomes negligible for very high impurity concentrations. This shows that the Usadel approximation does require very short scattering lengths in order to be able to neglect odd-parity correlations in the ferromagnet. 
Right next to the interfaces, to suppress $|f_p|$ to about 1/10 of $|f_\epsilon|$ requires a scattering length as short as $l=0.01\ \xi_0$. 
One needs to carefully estimate if this is then already
on the inter-atomic length scale, a length scale outside the scope of quasiclassical theory (both Eilenberger-Larkin-Ovchinnikov and Usadel theory), in which case the boundary layer must be treated microscopically. 
One option is to incorporate this region as an 'isotropization' zone in effective boundary conditions for the Usadel equation. 

We next discuss the Josephson effect in the structure shown in Fig.~\ref{sketch}.
With regard to the CPR, we first observe that higher harmonics, i.e. a non-sinusoidal CPR, can still be present in the high-transparency case, even if the suppression of pairing-correlations in the SC is properly taken into account. To show this, we consider a junction with high transparency interfaces in Fig.~\ref{impurities}. The CPR is calculated as a function of the impurity concentration. Decreasing the scattering length leads to a suppression of the critical current, but also of higher harmonics contributions, as can be seen in the normalized plots on the right-hand side.

\begin{figure}[t]
\includegraphics[width=\columnwidth]{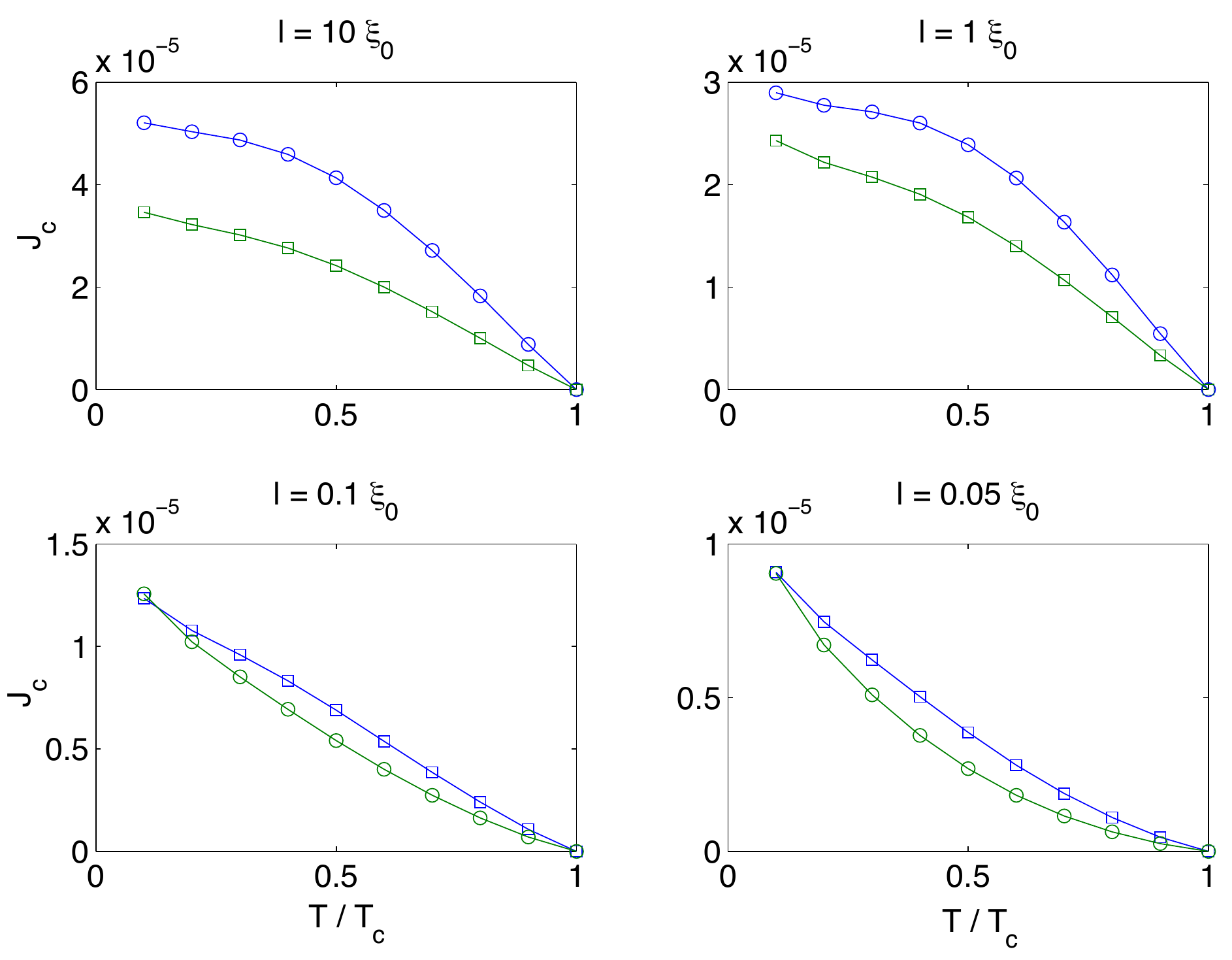}
\caption{\label{disorderT} Critical current in spin-up (blue, squares) and spin-down (green, circles) band for different impurity concentrations and $L=0.5\ \xi_0$ as a function of temperature. The low transparency interface is considered, where the critical current is found at $\Delta\chi=3/2\pi$. The current density is stated in units of $e k_{\rm B} \Tc (\pF/2\pi)^2/\hbar^3$. }
\end{figure}
In Fig.~\ref{disorderT}, we consider a junction in the tunneling limit ($d=2\ \lambda_{\rm F}/2\pi$, $V_I=2\ E_{\rm F}$) and calculate the critical current as a function of temperature. For a clean junction, $l=10\ \xi_0$, we find a strong spin-polarization of the current. For higher impurity concentrations, the current is not only suppressed globally, but the spin current, $I_\uparrow-I_\downarrow$, also decreases relative to the total current. At high impurity concentrations, there may even be a crossover to a negative spin-current at small temperatures. This can be understood as follows. From the clean-limit solution, we see that the loss of particle-hole coherence due to temperature penalizes the spin-down current more than the spin-up current. The mean free path due to impurity scattering is, however, the same for both bands, and the cumulative effect of the scattering only depends on the length of the trajectory. This can easily be understood by rewriting the transport equation as:
\begin{equation} (i\hbar\hat{\vec{v}}_{\rm F\eta}\nabla+i2\epsilon_n/v_{\rm F\eta})\gamma=[
\gamma \tilde \Delta \gamma+\Sigma \gamma-\gamma \tilde \Sigma-\Delta]/v_{\rm F\eta}.\end{equation}
Since all self-energies are proportional to $1/\tau_\eta\propto v_{\rm F\eta}$ (the superconducting gap is zero in the FM), the $v_{\rm F\eta}$ factor cancels on the right hand side of this equation, which describes impurity scattering, and only remains in $i\epsilon_n/v_{\rm F\eta}$ -- which is the term that can be interpreted as the suppression due to finite temperatures, and which clearly is seen to be spin-dependent.

\section{Conclusions}

We carried out a comprehensive study of the triplet-Josephson effect and inverse proximity effect in SC/FM/SC junctions with highly spin-polarized FM-layers taking into account impurity scattering inside the FM in the self-consistent Born-approximation. We showed that odd-momentum triplet correlations are generically present in these junctions and are only suppressed at very high impurity concentrations. We found an induced magnetization in the SC-electrodes resulting from the inverse proximity effect and a suppression of the SC-gap which is mainly related to the transparency of the interface for the interface model we considered here. Larger spin-mixing phases would, however, also result in a substantial suppression of the gap close to the interface \cite{eschrig08}. These can be realized by considering a smooth scattering potential at the interface. Even if this suppression is taken into account, the CPR of high-transparency junctions may have higher-harmonic contributions. Regarding the critical current, we see that impurity scattering suppresses the Josephson spin-current and may even lead to an inversion of it at small temperatures and short scattering lengths.

\acknowledgements

RG acknowledges financial support from the Karlsruhe House of Young Scientists. 
TL acknowledges support from the Swedish Research Council (VR).
ME acknowledges support from EPSRC under grant reference EP/J010618/1.

\appendix

\section{Expressions for the Fermi-surface average}
\label{FSA}

For the spherical bands that we assume, the density of states at the Fermi-level is:
\begin{align}
N_{\rm F,\eta}&=\int_{FS,\eta} \frac{{\rm d}^2 p_{\mathrm{F},\eta}}{(2\pi\hbar)^3|\vec{v}_{\mathrm{F},\eta}(\vec{p}_{\mathrm{F},\eta})|}=\frac{2m}{(2\pi)^2\hbar^3}\cdot p_{\rm F,\eta}.\end{align}
The FS-average for the self-energies is calculated from:
\begin{align}\label{FSself} \langle x(\vec{p}_{\rm F} )\rangle_\eta=\frac{1}{2 N_{\rm F, \eta}}\int_0^{p_{\rm F,\eta}} d\left[{\rm arcsin}\frac{p_{||}}{p_{\rm F,\eta}}\right]\ p_{||}\left(x^+_\eta+x^-_\eta\right),\end{align}
with $x^+_\eta=x_\eta(p_{||},p_z>0)$ and $x^-_\eta=x_\eta(p_{||},p_z<0)$. And the FS-average for the current:
\begin{align} \langle v_{\rm F,\eta} g\rangle_{\eta,+}=\frac{2\pi\cdot \hat{e}_z}{(2\pi\hbar)^3N_{\rm F,\eta}}\int_0^{p_{\rm F,\eta}}dp_{||}\ p_{||}\cdot \mathrm{Re}[g^+]. \end{align}
The number of $p$-points we use in our calculation (one hundred) is to small to make these expressions converge numerically. Thus, we use a numerical normalization factor $N_{\rm F,\eta}^{\rm num}$ which is calculated from Eq. \eqref{FSself} as $N_{\rm F,\eta}^{\rm num}=N_{\rm F,\eta}\cdot \langle 1\rangle_\eta$. This ensures that all our FS-averages are properly normalized. To be precise, the numerical integration is implemented as a trapezoidal rule over one hundred $p$-points which are equidistant in $p_{||}$. 

\section{Numerical iteration of gap and boundary conditions}
\label{NUM}
\subsection{Discretization of the gap-profile}
\label{GAP}

\begin{figure}
 \includegraphics[width=\columnwidth]{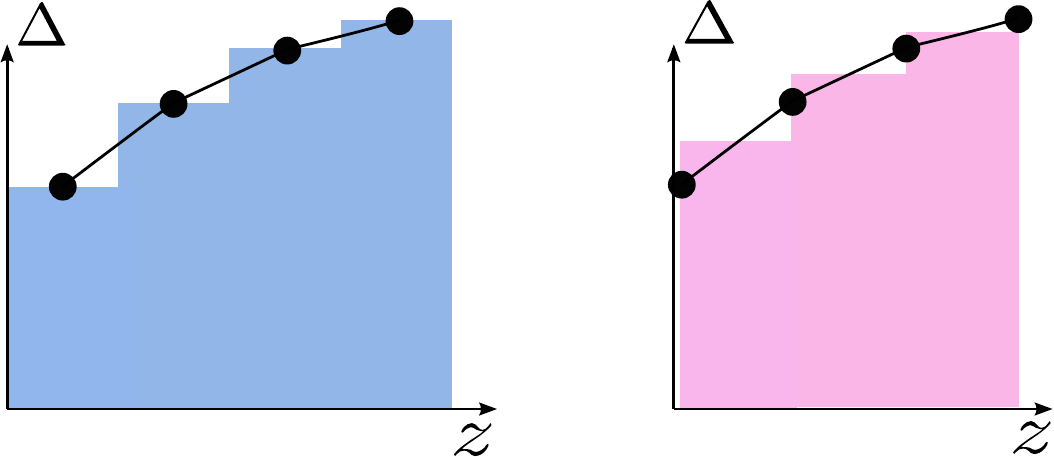}
 \caption{\label{profile}The discretized profile of the energy gap must be carefully chosen to ensure the consistency of the asymptotic solution of the transport equation. Above, the profile on the right-hand side, interpolating the profile as $1/2[\Delta(z_{n})+\Delta(z_{n+1})]$ between neighboring grid points violates this consistency, while the one on the left-hand side, that we use for our calculation, does not. }
 \end{figure}

The numerical solver we use is not a standard Runge-Kutta scheme but relies on the analytical solution for constant self-energies. The only numerical approximation made is assuming a step-wise variation of the self-energies between grid-points. Naturally, the question arises how this stepwise profile is to be interpolated when knowing the self-energies at a given number of grid-points. A subtlety here is that the coupled system of transport and self-consistency equations becomes fundamentally inconsistent if the condition $\Delta_{\rm interpol}(z=n\cdot h)= \Delta_n$, where $n$ denotes the $n$th grid-point, is violated. This is because the asymptotic solution of the transport equation approaches the local homogeneous bulk-solution as the Matsubara frequency goes to infinity. As we will see in appendix \ref{elimination}, this is crucial for formulating the gap-equation in a cut-off independent way.
The interpolated profile we use is illustrated in Fig.~\ref{profile} (left).

\subsection{Iterative scheme}
\label{IS}
We assume that at the beginning of the iteration step, the incoming solutions at all interfaces and the self-energies are known from the previous step. The procedure works as follows:
\begin{enumerate}
\item Solve the boundary conditions at all interfaces.
\item Integrate all outgoing coherence functions up to the next interface or outer boundary.
\item Update all self-energies. We use the secant method to accelerate convergence.
\item Update all incoming coherence functions.
\end{enumerate}
We use two abortion criteria, since the iteration in the SCs may converge a lot faster than in the central FM layer if the transparency of the interfaces is small. If the gap-profile is well-converged, the iteration in the SC is stopped and continues only inside the FM.

\subsection{Elimination of the high-energy cut-off}
\label{elimination}
Generally, technical energy cut-offs that appear in expressions for self-energies or observables in quasiclassical theory must be eliminated in favor of phenomenological parameters. This is usually achieved by adding and subtracting terms in Matsubara sums or energy integrals that cancel the divergent behavior of these expressions and allow to formally extend the sum or integral to infinity \cite{Serene}.  The other half of such an ``added zero'' can typically be included in a phenomenological (measurable) parameter. While this procedure is standard for the weak-coupling gap equation in quasiclassical theory, it turns out to be problematic at interfaces. We are not aware of any discussion of this subtle issue in the literature before.
We here discuss a method for dealing with this issue. We start from the self-consistency equation:
\begin{align}\label{self} \Delta(\pF, \vec{R})=k_{\rm B}T \sum_{|\epsilon_n|<\epsilon_c}  \langle V(\pF, \pF')f(\pF',\vec{R},\epsilon_n)\rangle.\end{align}
In what follows, we limit the discussion to singlet, s-wave superconductivity, i.e. $V(\pF,\pF')=V_0$, but the argumentation can easily be generalized. The interaction constant can be eliminated by linearizing the above equation close to $T_c$ ($\Delta \rightarrow 0$). Since the interaction constant is a bulk-property, it is justified to consider the homogeneous bulk case, i.e.:
\begin{align} f(\epsilon_n)=\pi\frac{\Delta}{\sqrt{\epsilon_n^2+\Delta^2}}\stackrel{\epsilon_n\gg \Delta}{\approx} \pi \frac{\Delta}{|\epsilon_n|}.\end{align}
This equation also shows why a technical cut-off $|\epsilon_n|<\epsilon_c$ must be included in \eqref{self}, as the Matsubara sum diverges. This problem does not arise in the microscopic theory, where the asymptotic behavior of $F$ is given by $\Delta/\epsilon_n^2$.\cite{Schrieffer} The linearized equation reads:
\begin{align} \frac{1}{V_0}=k_{\rm B} T_c \pi \sum_{|\epsilon_n|<\epsilon_c} \frac{1}{|\epsilon_n|}. \end{align}
Given that $n_c=\frac{\epsilon_c}{2\pi k_{\rm B} T_c}+\frac{1}{2}$, we then 
obtain the well-known standard result in terms of the digamma-function $\psi$,
\begin{align}\label{V} \frac{1}{V_0}=\psi\left(\frac{1}{2}+\frac{\epsilon_c}{2\pi k_{\rm B} T_c}\right)-\psi\left(\frac{1}{2}\right).\end{align}
At this point, one can see that keeping a technical cut-off is, at least in the Matsubara case, not tenable. $n_c$ should of course be an integer number. Obviously one can choose $\epsilon_c$ so that $n_c$ is integer here, but $\epsilon_c$ must be the same for all temperatures and since the spacing of Matsubara frequencies is temperature dependent, it will generally not be commensurate with $\epsilon_c$. One can verify numerically that if such a cut-off is kept in $\eqref{self}$, the gap-profile one finds is the BCS-relation with a jigsaw-pattern superimposed to it, which vanishes as $\epsilon_c\rightarrow \infty$. 
The standard way of dealing with this problem is to
regularize the Matsubara sum. This is achieved by subtracting the asymptotic form of $f$ on both sides of \eqref{self}:
\begin{align}& \Delta \Big(\frac{1}{V_0}-k_{\rm B} T\pi \sum_{|\epsilon_n|<\epsilon_c} \frac{1}{|\epsilon_n|}\Big)\\\nonumber
&=k_{\rm B} T \sum_{|\epsilon_n|<\epsilon_c} \Big(\langle f(\pF',\vec{R},\epsilon_n)\rangle-\pi \frac{\Delta}{|\epsilon_n|}\Big).
\end{align}
Both sides of this equation can now be regularized, if the expression \eqref{V} for $V_0$ is taken into account. The cut-off can be taken to infinity, which yields
\begin{align}\label{reg} \Delta \ln\left(\frac{T}{T_c}\right)=k_{\rm B} T \sum_{\epsilon_n} \Big(\langle f(\pF',\vec{R},\epsilon_n)\rangle-\pi \frac{\Delta}{|\epsilon_n|}\Big).\end{align}
For the boundary value problem one may object that the counter term was adapted to the homogeneous bulk-case, but expanding the solution of the Riccati-equations in orders of $1/\epsilon_n$, one finds \cite{EschrigThesis}:
\begin{align} \gamma(\pF, \vec{R},\epsilon_n)&=-\frac{\Delta(\vec{R})}{2i\epsilon_n}-\frac{i\hbar (\vec{v}_{\rm F} \nabla) \Delta(\vec{R})}{(2\epsilon_n)^2}+\mathcal{O}(1/\epsilon_n^3),\\\nonumber
 \tilde\gamma(\pF, \vec{R},\epsilon_n)&=\frac{\tilde\Delta(\vec{R})}{2i\epsilon_n}-\frac{i\hbar (\vec{v}_{\rm F} \nabla) \tilde\Delta(\vec{R})}{(2\epsilon_n)^2}+\mathcal{O}(1/\epsilon_n^3),\end{align}
 and therefore:
 \begin{align} f&=-2i\pi \sum_n (\gamma\tilde\gamma)^n\gamma\\\nonumber
 &=\pi\frac{\Delta(\vec{R})}{|\epsilon_n|}\mp2\pi\frac{\hbar (\vec{v}_{\rm F} \nabla) \Delta(\vec{R})}{(2\epsilon_n)^2}+\mathcal{O}(1/\epsilon_n^3),\ \epsilon_n \gtrless 0.\end{align}
 Thus, even if the gap-profile is inhomogeneous, the asymptotic behavior remains the same as long as the gap is a smoothly varying function.
 
Up to this point, the procedure is well-known and established. A problem does, however, arise at interfaces where the boundary conditions leads to a discontinuous behavior of the coherence functions under reflection. To see this, we consider the coherence function to leading order in $1/\epsilon_n$. The incoming coherence function at the interface is then given by $\gamma(z=0)=-\Delta(z=0)/(2i\epsilon_n)$. To illustrate the point we here assume that the solution of the boundary conditions at the interface is:
 \begin{align}\label{boundhigh} \Gamma=R_{\rm SC} \gamma(z=0)\tilde R_{\rm SC},\end{align}
where $R_{\rm SC}$ denotes the reflection matrix on the SC-side of the interface. This is not fully generic, since we assume that the transmission from the other side of the interface can be neglected. However, for the case we consider here this is reasonable as the coherence functions on the FM side are exponentially suppressed by propagation through the FM layer and can thus be neglected at high energies. Now it is easy to see that the asymptotic behavior of $\langle f\rangle$ at the interface is given by:
\begin{align}\nonumber \langle f(\pF,\epsilon_n)\rangle&=\langle  f(\pF,\epsilon_n)\rangle_++ f(\pF,\epsilon_n)\rangle_-\\
&=\frac{\pi}{2}\frac{\Delta}{|\epsilon_n|}\Big(1-i\sigma_y\langle R_{\rm SC}i\sigma_y\tilde R_{\rm SC}\rangle\Big).\end{align}
Here $\langle \bullet \rangle_\pm$ denotes an FS-average over trajectories pointing towards/away from the interface. It is obvious that this $f$-function will only have the required asymptotic behavior if $R_{\rm SC}=1$ or $\Delta =0$. This implies that the right-hand side of equation \eqref{reg} will diverge if that is not the case. Adapting the cancellation term will not help either, since the coupling constant $V_0$ is fixed by the bulk-behavior, which implies that the left-hand side of \eqref{reg} would diverge in that case. So eventually, one arrives at the conclusion that the only consistent solution for $R_{{\rm SC}}\ne 1$ is $\Delta(z=0)=0$. The problem here is the discontinuous jump of the $f$-function upon reflection at the interface, which is directly related to the fact that the interface region cannot be described within quasiclassical theory. While the use of effective boundary conditions is sufficient to circumvent this problem if the self-consistency relation is neglected, one must go a step further if it is retained.
Our solution to this problem is to assume that the self-consistency equation must not be calculated if the distance to the interface is smaller than a length cut-off $\xi_c \ll \xi_0$. To see how this resolves our issue, we have to calculate $\gamma(z=\xi_c)$ and $\Gamma(z=\xi_c)$. Here, $\gamma(z=\xi_c)$ is the incoming coherence function and obtained by integrating the Riccati equation from some initial value inside the bulk. Since the gap-profile is smooth for $z>\xi_c$, this implies that it does have the correct asymptotic behavior at high energies. $\Gamma(z=\xi_c)$ originates from the interface. If we assume that the gap has a constant value $\Delta_c=\Delta(z=\xi_c)$ in some environment $[\xi_c-\eta,\xi_c]$ of $\xi_c$, where $\eta$ can be arbitrarily small, the solution for $\Gamma(z=\xi_c)$ will be given by \cite{eschrig09}
\begin{align} \Gamma_c=\gamma_{h,c}+e^{i\eta\Omega_1}[\gamma_0-\gamma_{h,c}]\Big\{e^{i\eta\Omega_2}+C(\rho)\cdot[\gamma_0-\gamma_{h,c}]\Big\}^{-1},\end{align}
where $\gamma_{h,c}=\gamma_h(\Delta_c)$ and $\gamma_0=\Gamma(z=\xi_c-\eta)$. This solution has the correct asymptotic behavior, irrespective of how the gap-profile varies for $z<\xi_c-\eta$, as it is given by the homogeneous solution $\gamma_{h,c}$ at sufficiently high energies. 
We found that the sum in \eqref{reg} must be, however, calculated up to very high energies, since $\xi_c$ should be very short. Thus, we need to discuss high-energy contributions next.

\subsection{High-energy contribution}
\label{HEC}
To numerically compute \eqref{reg}, we consider two numerical energy cut-offs on the right-hand side:
\begin{align} \ln\left(\frac{T}{T_c}\right)\Delta&=\sum\limits_{|\epsilon_n|<\epsilon_{\rm c,low}}\Big(\langle f(\pF',\vec{R},\epsilon_n)\rangle-\pi \frac{\Delta}{|\epsilon_n|}\Big)\\\nonumber
&+\sum\limits_{\epsilon_{\rm c,high}>|\epsilon_n|\geq \epsilon_{\rm c,low}}\Big(\langle f(\pF',\vec{R},\epsilon_n)\rangle-\pi \frac{\Delta}{|\epsilon_n|}\Big).\end{align}
Up to $\epsilon_{\rm c,low}$ we calculate the full coherence functions in the whole structure without any approximations. The high-energy contribution is only calculated in the SC-electrodes. To efficiently calculate this contribution, we consider the linearized Riccati-equation:
\begin{align} i\partial_\rho \gamma(\rho)+2i\epsilon_n \gamma=-\Delta(\rho), \end{align}
which is easily solved exactly by a variation of constants:
\begin{align}\nonumber \gamma(\rho)&=[C(\rho)+\gamma_0]\beta(\rho),\ \beta=\exp[-2\epsilon_n\rho], \\C(\rho)&=\int_0^\rho d\rho' \frac{i\Delta(\rho')}{\beta(\rho')}. \end{align}
Again assuming that the gap-profile is constant, this yields:
\begin{align} \gamma(\rho)=\gamma_h+[\gamma_0-\gamma_h]\beta(\rho).\end{align}
We also assume that the boundary conditions reduce to \eqref{boundhigh}. 
It is easy to convince oneself that the linearized Riccati-equation is exact up to second order in $1/\epsilon_n$. Up to second order, we also have:
\begin{align} f=-2i\pi \gamma,\end{align}
from which we calculate the contribution to the gap function.

As discussed above, the need for taking into account this high-energy correction factor arises from interface scattering. It is therefore not surprising that it vanishes rapidly inside the SC-bulk. This fortunately implies that we must only calculate it very close to the interface.

\subsection{Extension to zero temperature}
\label{ZeroT}

For $T\rightarrow 0$, we have
\begin{align} &\epsilon_{n+1}-\epsilon_n=2\pi k_{\rm B} T \rightarrow 0,\\ &\ k_{\rm B}T\sum_{\epsilon_n} \rightarrow \frac{1}{2\pi} \int_{-\infty}^\infty d\epsilon.\end{align}
The zero-temperature Green function is thus obtained by replacing $\epsilon_n\mapsto \epsilon$ and all Matsubara sums must be replaced with the above integral. Care must be taken in regard to the normalization factor in the self-consistency equation:
\begin{align} \ln \left(\frac{T}{T_c}\right)+\sum_{n=1}^{n_c} \frac{1}{n-1/2}, \end{align}
with $n_c=\mathrm{floor}[\epsilon_c/(2\pi k_{\rm B}T)+1/2]$, since both terms in this sum become infinite. For $T\rightarrow 0$, the second term approaches $\ln[\epsilon_c/(2\pi k_{\rm B}T)]-\psi(1/2)$, and we have:
\begin{align} \ln \left(\frac{T}{T_c}\right)+\sum_{n=1}^{n_c} \frac{1}{n-1/2}=\ln\left(\frac{\epsilon_c}{2\pi k_{\rm B} T_c}\right)-\psi\left(\frac{1}{2}\right).\end{align}
\section{Calculation of the $S$-matrix}
\label{SM}

\subsection{HM-case}

First, we need the wave-vectors which are obtained from solving the one-dimensional Schr\"odinger-equation for given $k_x$, $k_y$. These are $k_+$ and $k_-$ for $H^{\rm I}$, $p$ for $H^{\rm SC}$, and $k_\u$ and $\kappa$ for $H^{\rm FM}$, where $\kappa$ is associated to the evanescent mode. These wave-vectors $k_\pm$ may be real or imaginary, corresponding to either a conducting or an insulating interface. Moreover, we have the Fermi-velocities $v$ and $v_\u$ in the SC and the majority band of the FM, respectively, which are proportional to $p$ and $k_\u$ respectively. After defining the auxiliary quantities
\begin{align} a_\pm&=\cos k_\pm a+\frac{ip}{k_\pm}\sin k_\pm a,\\
b_\pm&=\cos k_\pm a-\frac{i p}{k_\pm}\sin k_\pm a,\\
c_\pm&=\frac{ik_\pm}{k_\u}\sin k_\pm a+\frac{p}{k_\u}\cos k_\pm a,\\
d_\pm&=\frac{i k_\pm}{k_\u}\sin k_\pm a-\frac{p}{k_\u}\cos k_\pm a,
\end{align}
where $a$ is the thickness of the interface layer, and the $2\times 2$ matrices:
\begin{align} P(x)&=\mathcal{U}^\dagger \left(\begin{array}{cc} x_+ & 0 \\ 0 & x_- \end{array}\right) \mathcal{U},\end{align}
where $\mathcal{U}$ is the spin-rotation from the spin-basis of the interface to that of the FM-bulk, we obtain:
\begin{align} S^{\rm R}=U^{-1}V \end{align}
with
\begin{align} U&=\left(\begin{array}{ccc} -\sqrt{\frac{v_\u}{v}}P_{11}(b) & -\sqrt{\frac{v_\u}{v}}P_{12}(b) & 1\\ -\sqrt{\frac{v_\u}{v}}P_{11}(d) & -\sqrt{\frac{v_\u}{v}}P_{12}(d) & 1\\
P_{21}(d)-\frac{i\kappa}{k_\u}P_{21}(d) & P_{22}(d)-\frac{i\kappa}{k_\u}P_{22}(b) & 0 \end{array}\right),\\
V&=\left(\begin{array}{ccc} \sqrt{\frac{v_\u}{v}}P_{11}(a) & \sqrt{\frac{v_\u}{v}}P_{12}(a) & -1\\ \sqrt{\frac{v_\u}{v}}P_{11}(c) & \sqrt{\frac{v_\u}{v}}P_{12}(c) & 1\\
\frac{i\kappa}{k_\u}P_{21}(a)-P_{21}(c) & \frac{i\kappa}{k_\u}P_{22}(a)-P_{22}(c) & 0 \end{array}\right).\end{align}

\subsection{FM case}
In the FM case, the minority band of the FM has a propagating mode. We write $k_\u$ and $k_\d$ for the wave-vectors of the plane-waves in the FM-bands and $v_\u$, $v_\d$ for the corresponding group velocities. We use the same definitions for $a_\pm$, $b_\pm$ and $P(x)$ as above, but redefine:
\begin{align} c_\pm=i k_\pm \sin k_\pm a+ p \cos k_\pm a,\\
d_\pm=i k_\pm \sin k_\pm a - p \cos k_\pm a, \end{align}
and introduce:
\begin{align} R(x)&=\left(\begin{array}{cc} \sqrt{\frac{v_\u}{v}} & 0 \\ 0 & \sqrt{\frac{v_\d}{v}}\end{array}\right) P(x),\\
\tilde R(x)&=\left(\begin{array}{cc} \sqrt{\frac{v_\u}{v}} & 0 \\ 0 & \sqrt{\frac{v_\d}{v}}\end{array}\right)\left(\begin{array}{cc} \frac{1}{q_+} & 0 \\ 0 & \frac{1}{q_-}\end{array}\right)P(x).\end{align}
We then redefine 
\begin{align} U=\left(\begin{array}{cc} -R(b)& 1_{2\times2} \\ -\tilde{R}(d) & 1_{2\times2} \end{array}\right),\ 
V=\left(\begin{array}{cc} R(a)& -1_{2\times2} \\ \tilde{R}(c) & 1_{2\times2} \end{array}\right) \end{align}
and again obtain the scattering matrix as:
\begin{align} S^{\rm R}=U^{-1}V. \end{align}


\vspace{-0.5cm}
\begin{thebibliography}{27}
\expandafter\ifx\csname natexlab\endcsname\relax\def\natexlab#1{#1}\fi
\expandafter\ifx\csname bibnamefont\endcsname\relax
  \def\bibnamefont#1{#1}\fi
\expandafter\ifx\csname bibfnamefont\endcsname\relax
  \def\bibfnamefont#1{#1}\fi
\expandafter\ifx\csname citenamefont\endcsname\relax
  \def\citenamefont#1{#1}\fi
\expandafter\ifx\csname url\endcsname\relax
  \def\url#1{\texttt{#1}}\fi
\expandafter\ifx\csname urlprefix\endcsname\relax\def\urlprefix{URL }\fi
\providecommand{\bibinfo}[2]{#2}
\providecommand{\eprint}[2][]{\url{#2}}

\bibitem{bergeret05}
F.~S. Bergeret, A.~F. Volkov, and K.~B. Efetov, Rev. Mod. Phys. {\bf 77}, 1321 (2005).

\bibitem[{\citenamefont{{A.~I. Buzdin}}(2005)}]{buzdin05}
\bibinfo{author}{\bibnamefont{{A.~I. Buzdin}}}, \bibinfo{journal}{Rev. Mod.
  Phys.} \textbf{\bibinfo{volume}{77}}, \bibinfo{pages}{935}
  (\bibinfo{year}{2005}).

\bibitem{eschrig07}
M. Eschrig, T. L\"ofwander, T. Champel, J.C. Cuevas, and G. Sch\"on, J. Low Temp. Phys. {\bf 147}, 457 (2007).

\bibitem{lofwander10}
T. L\"ofwander, R. Grein, and M. Eschrig, Phys. Rev. Lett. {\bf 105}, 207001 (2010);
S. Piano, R. Grein, C.J. Mellor, K. V\'yborn\'y, R. Campion, M. Wang, M. Eschrig, and B.L. Gallagher, Phys. Rev. B {\bf 83}, 081305(R) (2011).

\bibitem{eschrig11}
M. Eschrig, Physics Today, January 2011 Issue, p.43 (2011)

\bibitem{ryazanov01}
V. V. Ryazanov, V. A. Oboznov, A. Yu. Rusanov, A. V. Veretennikov, A. A. Golubov, and J. Aarts, Phys. Rev. Lett. {\bf 86}, 2427 (2001)

\bibitem{kontos02}
T. Kontos, M. Aprili, J. Lesueur, F. Gen\^ot, B. Stephanidis, and R. Boursier, Phys. Rev. Lett. {\bf 89}, 137007 (2002)

\bibitem{blum02}
Y. Blum, A. Tsukernik, M. Karpovski, and A. Palevski, Phys. Rev. Lett. {\bf 89}, 187004 (2002)

\bibitem{keizer06}
R.S. Keizer, S.T.B. Goennenwein, T.M. Klapwijk, G. Miao, G. Xiao, and A. Gupta,
Nature (London) {\bf 439}, 825 (2006).

\bibitem{sosnin06}
I. Sosnin, H. Cho, V. T. Petrashov, and A. F. Volkov, Phys. Rev. Lett. {\bf 96}, 157002 (2006)

\bibitem{robinson10a}
J. W. A. Robinson, J. D. S. Witt, and M. G. Blamire, Science {\bf 329}, 59 (2010)

\bibitem{robinson10b}
J. W. A. Robinson G. B. Halasz, A. I. Buzdin, and M. G. Blamire, Phys. Rev. Lett. {\bf 104}, 207001 (2010)

\bibitem{khaire10}
T. S. Khaire, M. A. Khasawneh, W. P. Jr. Pratt, and N. O. Birge, Phys. Rev. Lett. {\bf 104}, 137002 (2010)

\bibitem{anwar10}
M. S. Anwar, F. Czeschka, M. Hesselberth, M. Porcu, and J. Aarts, Phys. Rev. B {\bf 82}, 100501(R) (2010)

\bibitem{sprungmann10}
D. Sprungmann, K. Westerholt, H. Zabel, M. Weides, and H. Kohlstedt, Phys. Rev. B {\bf 82}, 060505(R) (2010)

\bibitem{khasawneh11}
M. A. Khasawneh, T. S. Khaire, C. Klose, W. P. Jr. Pratt, N. O. Birge, Supercond. Sci. Technol. {\bf 24}, 024005 (2011)

\bibitem{bergeret01}
F. S. Bergeret, A. F. Volkov, and K. B. Efetov, Phys. Rev. Lett. {\bf 86}, 4096 (2001)

\bibitem{volkov03}
A. F. Volkov, F. S. Bergeret, and K. B. Efetov, Phys. Rev. Lett. {\bf 90}, 117006 (2003)

\bibitem{eschrig03}
M. Eschrig, J. Kopu, J. C Cuevas, and G. Sch\"on, Phys. Rev. Lett. {\bf 90},
137003 (2003).

\bibitem{houzet07}
M. Houzet, and A. I. Buzdin, Phys. Rev. B {\bf 76}, 060504 (2007)

\bibitem{eschrig08}
M. Eschrig and T. L\"ofwander, Nat. Phys. {\bf 4}, 138 (2008).

\bibitem{grein09}
R. Grein, M. Eschrig, G. Metalidis, and G. Sch\"on,
Phys. Rev. Lett. {\bf 102}, 227005 (2009).

\bibitem{eschrig09}
M. Eschrig, Phys. Rev. B {\bf 80}, 134511 (2009).

\bibitem{grein10}
R. Grein, T. L\"ofwander, G. Metalidis, and M. Eschrig, Phys. Rev. B {\bf 81}, 094508 (2010)

\bibitem{usadel70}
K. Usadel, Phys. Rev. Lett. {\bf 25}, 507 (1970)

\bibitem{gorkov58}
L.~P. Gor'kov, Zh. Eksp. Teor. Fiz. {\bf 34},  735  (1958), [Sov.\ Phys.\ JETP
  {\bf7}, 505 (1958)],
L.~P. Gor'kov, Zh. Eksp. Teor. Fiz. {\bf 36},  1918  (1959), [Sov.\ Phys.\ JETP
  {\bf9}, 1364 (1959)].

\bibitem{eilen}
G. Eilenberger, Z. Phys. {\bf 214},  195  (1968).

\bibitem{larkin68}
A.~I. Larkin and Y.~N. Ovchinnikov, Zh. Eksp. Teor. Fiz. {\bf 55},  2262
  (1968), [Sov. Phys. JETP {\bf28}, 1200 (1969)].

\bibitem{Serene}
J.~W. Serene and D. Rainer, Phys. Rep. {\bf 101},  221  (1983).

\bibitem{rammer86}
J. Rammer and H. Smith, Rev. Mod. Phys. {\bf 58}, 323 (1986).

\bibitem{matsubara55}
T. Matsubara, Prog. Theor. Phys. (Kyoto) {\bf 14}, 351 (1955)

\bibitem{schopohl95}
N. Schopohl and K. Maki, Phys. Rev. B {\bf 52},  490  (1995),
N. Schopohl, cond-mat/9804064 (unpublished, 1998).

\bibitem{EschrigThesis} M. Eschrig, PhD thesis, University of Bayreuth (1997).

\bibitem{eschrig99}
M. Eschrig, J. A. Sauls, and D. Rainer, Phys. Rev. B {\bf 60}, 10447 (1999).

\bibitem{eschrig00}
M. Eschrig, Phys. Rev. B {\bf 61}, 9061 (2000).

\bibitem{Schrieffer}
J. R. Schrieffer, \textit{Theory of Superconductivity}, Perseus Books (1964).

\end{thebibliography}
\end{document}